\documentclass[a4paper,12pt]{article}
\usepackage{graphicx}
\usepackage{amssymb}
\usepackage{amsmath}

\newcommand{\mer}{m_{{\tilde{e}_R}}}
\newcommand{\mel}{m_{{\tilde{e}_L}}}
\newcommand{\msnu}{m_{{\tilde{\nu}}}}

\newcommand{\smur}{\tilde{\mu}_R}
\newcommand{\snu}{\tilde{\nu}_L}
\newcommand{\snue}{\tilde{\nu}_e}
\newcommand{\snut}{\tilde{\nu}_{\tau}}

\newcommand{\sul}{\tilde{u}_L}
\newcommand{\sdl}{\tilde{d}_L}
\newcommand{\stau}{\tilde{\tau}}
\newcommand{\staul}{\tilde{\tau}_1}

\newcommand{\mupl}{m_{{\tilde{u}_L}}}
\newcommand{\mupr}{m_{{\tilde{u}_R}}}
\newcommand{\msql}{m_{{\tilde{q}_L}}}
\newcommand{\mdnl}{m_{{\tilde{d}_L}}}
\newcommand{\mdnr}{m_{{\tilde{d}_R}}}
\newcommand{\mlsp}{m_{{\tilde{\chi}^0_1}}}
\newcommand{\mzii}{m_{{\tilde{\chi}^0_2}}}
\newcommand{\mziii}{m_{{\tilde{\chi}^0_3}}}
\newcommand{\mziv}{m_{{\tilde{\chi}^0_4}}}
\newcommand{\mtaul}{m_{\tilde{\tau}_1}}
\newcommand{\mtauh}{m_{\tilde{\tau}_2}}
\newcommand{\mntau}{m_{\tilde{\nu}_{\tau}}}
\newcommand{\mbl}{m_{\tilde{b}_1}}
\newcommand{\mbh}{m_{\tilde{b}_2}}
\newcommand{\mtl}{m_{\tilde{t}_1}}
\newcommand{\mth}{m_{\tilde{t}_2}}
\newcommand{\ziv}{\tilde{\chi}^0_4}
\newcommand{\ziii}{\tilde{\chi}^0_3}
\newcommand{\wii}{\tilde{\chi}^+_2}
\newcommand{\wi}{\tilde{\chi}^+_1}
\newcommand{\mwi}{m_{\tilde{\chi}^+_1}}
\newcommand{\mwii}{m_{\tilde{\chi}^+_2}}
\newcommand{\lsp}{\tilde{\chi}^0_1}
\newcommand{\zii}{\tilde{\chi}^0_2}
\newcommand{\sq}{\tilde{q}}
\newcommand{\psla}{p\kern-.45em/}
\newcommand{\esla}{E\kern-.45em/}

\newcommand{\beq}{ \begin{eqnarray} }
\newcommand{\eeq}{ \end{eqnarray} }
\newcommand{\beqstar}{ \begin{eqnarray*} }
\newcommand{\eeqstar}{ \end{eqnarray*} }
\newcommand{\gsim}{ \mathop{}_{\textstyle \sim}^{\textstyle >} }
\newcommand{\lsim}{ \mathop{}_{\textstyle \sim}^{\textstyle <} }

\newcommand{\cha}{\tilde{\chi}^+}

\newcommand{\neu}{\tilde{\chi}^0}

\newcommand{\sel}{\tilde{e}_L}
\newcommand{\ser}{\tilde{e}_R}

\newcommand{\sqb}{\tilde{q}^\star}
\newcommand{\sti}{\tilde{t}_1}
\newcommand{\sbi}{\tilde{b}_1}
\newcommand{\sbii}{\tilde{b}_2}
\newcommand{\gl}{\tilde{g}}
\newcommand{\mgl}{m_{\tilde{g}}}

\newcommand{\msti}{m_{\tilde{t}_1}}

\newcommand{\tchi}{\tilde{\chi}}
\newcommand{\xax}{x_{AX}}
\begin{document}
\thispagestyle{empty}
\vspace*{-15mm}
%----------
\baselineskip 10pt
\begin{flushright}
\begin{tabular}{l}
{\bf KEK-TH-800}\\
{\bf YITP-02-8}\\
{\bf hep-ph/0202129}
\end{tabular}
\end{flushright}
\baselineskip 24pt 
\vglue 10mm 
%%%%%%%%%%%%%%%%%%%%%%%%%%%%%%%%%%%%%%%%%%%%%%%%%%%%%%%%%%
%%%%%%%%%       TITLE      %%%%%%%%%%%%%%%%%%%%%%%%%%%%%%%
%%%%%%%%%%%%%%%%%%%%%%%%%%%%%%%%%%%%%%%%%%%%%%%%%%%%%%%%%%
\begin{center}
{\Large\bf
Slepton Oscillation\\
 at Large Hadron Collider
}
\\
\vspace{8mm}

\baselineskip 18pt 
\def\thefootnote{\fnsymbol{footnote}}
\setcounter{footnote}{0}

{\bf Junji Hisano$^{1)}$, Ryuichiro Kitano$^{1,2)}$ and Mihoko M.~Nojiri$^{3)}$}
\vspace{5mm}

$^{1)}${\it 
Theory Group, KEK,  Tsukuba, Ibaraki 305-0801, Japan}\\
$^{2)}${\it 
Department of Particle and Nuclear Physics,} \\
{\it The Graduate University for Advanced 
Studies,}\\
{\it  Tsukuba, Ibaraki 305-0801, Japan}\\
$^{3)}${\it 
YITP, Kyoto University, Kyoto 606-8502, Japan}
\vspace{10mm}
\end{center}
%%%%%%%%%%%%%%%%%%%%%%%%%%%%%%%%%%%%%%%%
%%%%%                              %%%%%
%%%%%          Abstract            %%%%%
%%%%%                              %%%%%
%%%%%%%%%%%%%%%%%%%%%%%%%%%%%%%%%%%%%%%%
\begin{center}
{\bf Abstract}\\[7mm]
\begin{minipage}{14cm}
\baselineskip 16pt
\noindent
Measurement of Lepton-Flavor Violation (LFV) in the minimal SUSY
Standard Model (MSSM) at Large Hadron Collider (LHC) is studied
based on a realistic simulation. We consider the LFV decay of the
second-lightest neutralino, $\tilde{\chi}^0_2 \to \tilde{l} l' \to l
l' \tilde{\chi}^0_1$, in the case where the flavor mixing exists in the
right-handed sleptons.  We scan the parameter space of the minimal
supergravity model (MSUGRA) and a more generic model in which we take
the Higgsino mass $\mu$ as a free parameter.  We find that the
possibility of observing LFV at LHC is higher if $\mu$ is smaller than
the MSUGRA prediction; the LFV search at LHC can cover the parameter
range where the $\mu \to e \gamma$ decay can be suppressed by the
cancellation among the diagrams for this case.

%%%%%%%%%% %%%%% %%%%% %%%%% %%%%% %%%%% %%%%% %%%%% %%%%% %%%%% %%%%% 
%%%%%-----------------------------------------------------------------
\end{minipage}
\end{center}
\newpage
\baselineskip 18pt 

\section{Introduction} 

The minimal supersymmetric (SUSY) standard model (MSSM) is one of the
attractive extensions of the standard model (SM). Many experiments are
searching for the possible evidence of the low-energy
supersymmetry. Among those, Lepton-Flavor Violation (LFV) processes may be
considered as major discovery modes of supersymmetry; they do
not exist in the SM or very small even if the small
neutrino masses are introduced.

In the MSSM the off-diagonal components of the slepton mass terms
violate lepton-flavor conservation, and they are related to the origin
of the SUSY breaking terms and interactions in physics beyond the
MSSM. An approximated universality of the sfermion masses should be
imposed at some energy scale so that the flavor-changing processes are
suppressed below the experimental bounds. One of the candidates to
realize the universality is the minimal supergravity (MSUGRA) model
\cite{Chamseddine:1982jx}. In this model, the LFV slepton masses are
induced by interactions above the GUT scale \cite{Hall:1985dx} or the
right-handed neutrino scale \cite{Borzumati:1986qx}. It is desirable
to discover LFV in different processes so that we can reconstruct the
off-diagonal slepton mass parameters which probe such interactions at
the higher energy scale.

Main constraints for the off-diagonal components of the slepton mass
matrices come from the rare decay process searches at low energy, such as
$\mu\rightarrow e \gamma$, $\mu N\rightarrow e N$, and
$\tau\rightarrow \mu \gamma$. The current experimental limits are
\begin{eqnarray}
Br(\mu\rightarrow e\gamma)&< 1.2\times 10^{-11}&\cite{Brooks:1999pu},
\cr
Br(\mu N\rightarrow eN)&<  6.1\times 10^{-13}&\cite{mueconv},
\cr
Br(\tau\rightarrow \mu\gamma)&< 1.0\times 10^{-6}&\cite{belltau}.
\end{eqnarray}
Once the approximate universality of the scalar masses is imposed,
the branching ratios are suppressed due to the GIM mechanism, and they
can be smaller than these experimental bounds for reasonable SUSY
parameter space. Therefore, those limits are not too serious at
present. In future, some of the proposed experiments aim to reach to
\begin{eqnarray}
Br(\mu\rightarrow e\gamma)&\lsim 10^{-14}&\cite{PSI},\cr
Br(\mu N\rightarrow e N) &\lsim 10^{-16}&\cite{meco}, \cr
		         &\lsim 10^{-18}&\cite{PRISM,nf}, \cr
Br(\tau \rightarrow \mu \gamma)&\lsim 10^{-(7-8)}&\cite{ohshima}.
\nonumber
\end{eqnarray}

LFV may be searched for at future collider experiments through the
production and decay of the slepton. While the rare lepton decay widths
suffer the suppression of the order of $(\Delta
m_{\tilde{l}}/\overline{m}_{\tilde l})^2$, the production and decay
processes of the slepton merely receive the suppression of the order of
$(\Delta m_{\tilde{l}}/\Gamma_{\tilde{l}})^2$
\cite{Arkani-Hamed:1996au}. Here, $\Delta m_{\tilde{l}}$ is the mass
difference between the sleptons, and $\overline{m}_{\tilde{l}}$ and
$\Gamma_{\tilde{l}}$ are the average mass and decay width of the
sleptons, respectively.  Thus, the future high-energy collider
experiments could explore the region of the parameter space which may not
be reached to by the rare decay searches. At future $e^+e^-$ linear
collider experiments, the $\tilde{e}$ production cross section could
be very large if the bino-like neutralino mass $M_1$ is relatively
light, and then the $e$-$\mu( \tau )$ mixing may be discovered there
\cite{Arkani-Hamed:1996au}\cite{Hirouchi:1997cy}.  Similarly, a muon
collider has a potential to access to $\mu$-$\tau$ mixing
\cite{Hisano:1998wn}.
% to fix

For the LFV search at Large Hadron Collider (LHC), the useful mode is
the LFV decay of the second-lightest neutralino ($\tilde{\chi}^0_2$),
$\tilde{\chi}^0_2 \to \tilde{l} l \to \tilde{\chi}^0_1 l l'$, where
the slepton oscillation effect leads to the different flavors for the
leptons in the final state ($l$ and $l'$).  This mode has an advantage
to observe LFV compared with the direct Drell-Yang production of the
sleptons, since $\tilde{\chi}^0_2$ can be copiously produced through
the cascade decay of the squarks and gluinos \cite{Agashe:1999bm}.
Typically 60\% of the first- and second-generation left-handed squark
decays into the wino-like neutralino and chargino, and in various
models the right-handed slepton masses are predicted to be smaller
than the second-lightest neutralino mass.  By using the above mode,
LFV in $\tilde{e}$-$\tilde{\mu}$ mixing has been investigated by
Agashe and Graesser \cite{Agashe:1999bm}, in which they chose the
point 5 of ATLAS TDR study \cite{TDR} in the MSUGRA model.  The
$\tilde{\tau}$-$\tilde{\mu}$ mixing has been recently studied by
Hinchliffe and Paige \cite{Hinchliffe:2001np}.

In this paper, we estimate the reach of LFV at LHC based on a
realistic simulation in the MSUGRA as well as a more generic model.
We assume that LFV comes from mixing of the right-handed smuon and
selectron.
The signal of LFV $\tilde{\chi}^0_2$ decay 
is the two opposite-sign leptons 
($e^+ \mu^-$ or $e^- \mu^+$) in the final state.
The distribution of the lepton-pair invariant mass
($m_{e\mu}$) in the LFV final state has an edge 
whose value is known because it is the same as
the edge of the lepton-flavor conserving opposite-sign modes
($m_{ee}$ and $m_{\mu\mu}$).
Since the distribution of the background processes does not have an
edge, we can obtain a sizable $S/N$ ratio in the region near the edge
of $m_{e \mu}$.

In the MSUGRA model, LHC can reach to $M\sim 400$ GeV where $M$ is the
common gaugino mass, when the mass difference and mixing angle of the
right-handed selectron and smuon are $\Delta m=1.2$ GeV and $\sin
2\theta= 0.5$. Unfortunately, the broad parameter space in the MSUGRA,
which can be probed by LHC, are already excluded by $\mu\rightarrow e
\gamma$ constraint. We point out that this is only for a case $\mu\gg
M$ where $\mu$ is the Higgsino mass.  The $\mu$ parameter may be close
to the gaugino masses in the more generic model, and in the case the
reduction of $Br(\mu\rightarrow e \gamma)$ by cancellation among the
diagrams is generic. We study a model of this kind (cMSSM) and we find
that the experimental reach of LFV search at LHC extends up to $M\sim
500$ GeV for the same slepton oscillation parameter.

We organize this article as follows. In the next section we discuss
prospect of the LFV searches at LHC in the MSUGRA and the cMSSM, and
present sensitivity at LHC for generic parameters in Section 3.
Section 4 is devoted to conclusions and discussion. In Appendix A we
present formula we used in this paper, and in Appendix B the mass
spectrum and the branching ratios for the SUSY particles are given for
the sample parameter sets we adopt in this paper.

\section{Prospect of  LFV searches at LHC in the MSUGRA and the more 
generic models }  

In the MSUGRA model, the GUT scale scalar mass $m$ and triliniear
coupling $A$ are universal among generations, and the gaugino masses
in the MSSM are given by the SU(5) gaugino mass $M$. The universal
scalar mass is predicted when the SUSY breaking sector couples to
chiral multiplets equally. However, it is possible, and might be
natural, to assume the different scalar masses between the matter
fields and the Higgs fields in the GUT context.
For example, in the most simple SO(10) model, we can set the following
boundary condition,
\begin{eqnarray}
&m_{\tilde{Q}}=m_{\tilde{q}_R} = m_{\tilde{l}_L}= m_{\tilde{l}_R}(\equiv
m_{16}),&\cr
&m_{H_1}= m_{H_2}(\equiv m_{10}).&
\label{soten}
\end{eqnarray}
We call this choice as the cMSSM. The MSUGRA model predicts $\mu\gg M$. On
the other hand, the cMSSM allows $\mu$ comparable to $M$.  We will
discuss the phenomenology of these models in the context of the LFV
search at LHC.

\subsection{The MSUGRA model}

We summarize qualitative features of the MSUGRA predictions relevant to 
the LFV study at LHC.

\begin{itemize}
\item $\mu\gg M$\\ 
The Higgs masses at the GUT scale are common to the other scalar
masses. The Higgsino mass parameter $\mu$ can be calculated as a
function of $m$, $M$, and $\tan\beta$ so as to reproduce the correct
electroweak symmetry breaking.  Especially for $m_t\sim 175$ GeV, the
dependence on $m$ effectively disappears when $m$ is of the order of
$M$, which is sometimes referred to the focus point
\cite{Feng:1999mn}.  In such a case, $\mu/M$ is more or less fixed,
and the LSP ($\tilde{\chi}^0_1$), the lighter chargino
($\tilde{\chi}^+_1$), and the second-lightest neutralino
($\tilde{\chi}^0_2$) are gaugino-like. The $\mu$ parameter is lighter
than the gaugino masses only when $m\gg M$, while the decay of
$\tilde{\chi}^0_2$ to $\tilde{l}$, which is relevant to our study, is
closed in the case. Thus, one can safely assume $\tilde{\chi}^0_2$ and
$\tilde{\chi}^0_1$ are wino- or bino-like in our study.

\item Light $\tilde{\tau}_1$\\
The lighter stau  $\tilde{\tau}_1$ is  lighter than the
other sleptons due to the left-right mixing proportional to
$\tan\beta\times \mu$. Note that the lower limit of the Higgs mass now
strongly constrain small $\tan\beta$ cases \cite{higgslimit}, and therefore
the mixing is expected to be significant. The typical outcome is
an increased branching ratio of $\tilde{\chi}^0_2$ to $\tilde{\tau}$, and
reduced branching fractions to the first- and the second-generation
sleptons $Br(\tilde{\chi}^0_2 \rightarrow \tilde{e},\tilde{\mu})$.
\end{itemize}

\begin{figure}[htbp]
\begin{center}
\includegraphics[width=7.5cm,angle=-90]{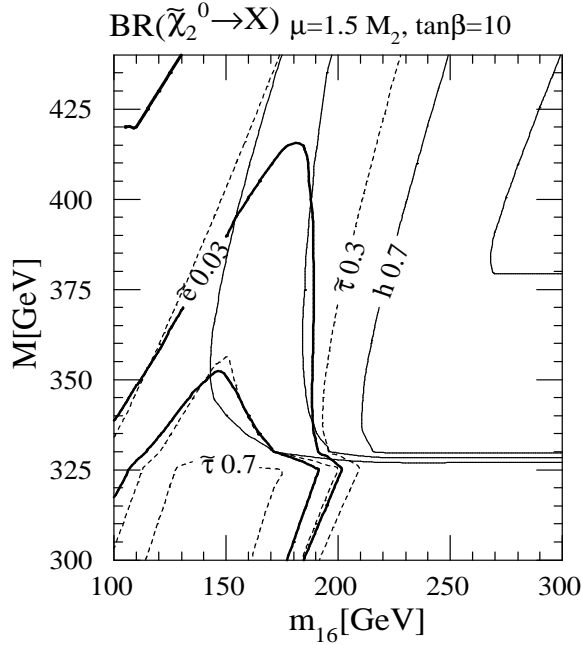}
\end{center}  
\caption{\footnotesize Contours of the constant branching ratios in the cMSSM
model for $\tan\beta=10$ and $\mu=1.5 M_2$. Solid lines are for
$Br(\tilde{\chi}^0_2\rightarrow h \tilde{\chi}^0_1)$ = 0.3, 0.5, 0.7,
0.9 (from left to right), dotted lines for
$Br(\tilde{\chi}^0_2\rightarrow \tau \tilde{\tau}_1)=$ 0.7, 0.5
0.3 (from bottom left), and thick solid lines
for $Br(\tilde{\chi}^0_2\rightarrow e \tilde{e}_R)=$ 0.05, 0.03 (from
bottom left). No solution above the thick line at the top left.  
\label{br_mu15m2.eps}}
\end{figure}

The fact that the MSUGRA model tends to reduce the branching ratio of
$\tilde{\chi}^0_2$ to the first- and second-generation sfermions
directly limits potential to search for LFV at LHC. This can be seen
in Fig.~\ref{br_mu15m2.eps}, where we plot contours of some constant
branching ratios of $\tilde{\chi}^0_2$ for $\tan\beta=10$ and $\mu=
1.5 M_2$ in the cMSSM, where $M_2$ is the gaugino mass of SU(2)$_L$.
The $\mu/M_2$ ratio is roughly what the MSUGRA model predicts for
moderate $\tan\beta$ and $M\sim m$.  Due to the Higgs mass constraint,
we do not plot the region where $M<300$ GeV. It is found from this
figure that the branching ratios to $\tilde{e}_R$ or $\tilde{\mu}_R$
are less than 6\% for $M>420$GeV. The reasons are the
following. First, note that the decay into the Higgs boson opens for
$M>325$ GeV, and it quickly dominates over the other decay processes.
Since $\tilde{\chi}^0_2$ is wino-like, the decay
$\tilde{\chi}^0_2\rightarrow h \tilde{\chi}^0_1$ is suppressed by only
$N_H\propto 1/\mu$, while $\tilde{\chi}^0_2\rightarrow$
$\tilde{e}_{R}$ is suppressed by $N_B\propto 1/\mu^2$. Here $N_H$ and
$N_B$ is the Higgsino and the bino components of $\tilde{\chi}^0_2$,
respectively.  As $\mu$ is relatively high in the MSUGRA model, the
branching ratio to $h$ is significantly larger than that to
$\tilde{l}_R$. Second, large $\mu \tan \beta$ induces non-negligible
$\tilde{\tau}$ left-right mixing.  Due to the left-right mixing,
$\tilde{\chi}^0_2$ decays into $\tilde{\tau}_1$ through the
$\tilde{\tau}_L$ component in $\tilde{\tau}_1$, and the decay
branching ratio dominates over the branching ratios to $\tilde{e}_R$
or $\tilde{\mu}_R$.  Third, if $m \ll M$, the decay into $\tilde{l}_L$
is open, and this quickly dominates over the other decay modes.  Thus,
the region where the decay of $\zii$ to $\tilde{e}_R$ or
$\tilde{\mu}_R$ has a sizable branching ratio is limited. We will
discuss the LFV signal in the next section.

\subsection{The cMSSM with $\mu\sim M_2$ }

We now turn into the case where universality of the sfermion masses
does not hold for the Higgs sector as Eq.~(\ref{soten}). We first
argue that this model predicts different phenomenology for the LFV
search.

\begin{figure}[htbp]
\begin{center}
\includegraphics[width=7.5cm,angle=-90]{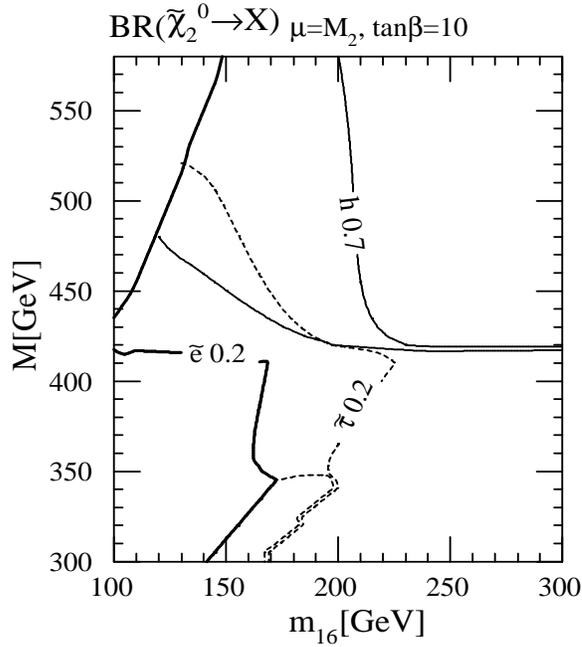}
\end{center}  
\caption{
\footnotesize 
Contours of constant branching ratios in the cMSSM
model for $\tan\beta=10$ and $\mu= M_2$. Solid lines are for
$Br(\tilde{\chi}^0_2\rightarrow h \tilde{\chi}^0_1)$ $= 0.5, 0.7$
(from left to right), dotted lines for $Br(\tilde{\chi}_2\rightarrow
\tau \tilde{\tau}^0_1)=$ $ 0.6, 0.2$ (from bottom left), and thick
solid lines for $Br(\tilde{\chi}^0_2\rightarrow e \tilde{e}_R)=$ $
0.2$. The LSP is charged above the thick solid line at top left. 
\label{br}}
\end{figure}

\begin{figure}[htbp]
\begin{center}
\includegraphics[width=7.5cm,angle=-90]{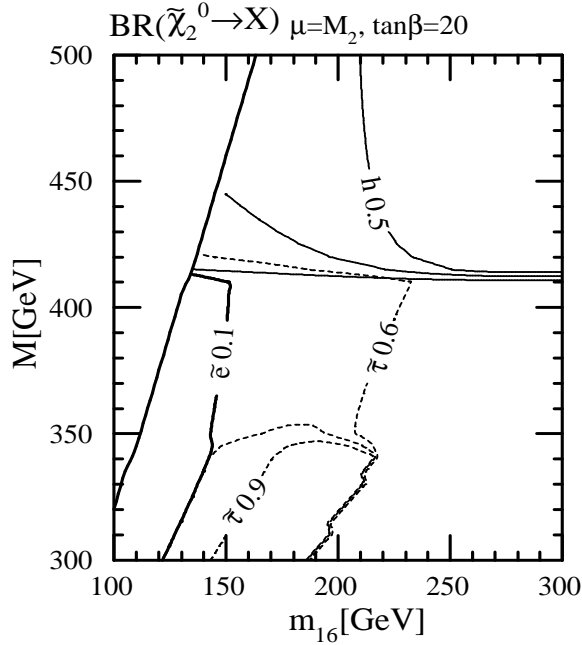}
\end{center}  
\caption{\footnotesize Contours of constant branching ratios in the cMSSM
model for $\tan\beta=20$ and $\mu= M_2$. Solid lines are for
$Br(\tilde{\chi}^0_2\rightarrow h \tilde{\chi}^0_1)$ $= 0.1, 0.3,0.5 $
(from bottom to top), dotted lines for $Br(\tilde{\chi}^0_2\rightarrow$ $
\tau \tilde{\tau}_1)=$ $ 0.9, 0.8, 0.6$ (from bottom to top), a
thick solid line for $Br(\tilde{\chi}^0_2\rightarrow e \tilde{e}_R)=$ $
0.1$. The LSP is charged above the thick solid line at left. 
\label{br_tanb20}}
\end{figure}

In Figs.~\ref{br} and \ref{br_tanb20}, we show the $\tilde{\chi}^0_2$
branching ratios for $\mu=M_2$ and $\tan\beta=10$, $20$.  Not only the
decay to $\tilde{e}_R$ is kinematically open, the branching ratio is
larger than that of the MSUGRA case. This is because the relatively
large Higgsino-gaugino mixing makes the $\tchi^0_2$ mass smaller than
$M_2$ and the mass difference between $\tchi^0_2$ and $\tchi^0_1$
becomes smaller. This makes the decay of $\tchi^0_2$ to $h$ close up
to $M\sim 420$ GeV.  At the same time, the decay into $\tilde{l}_L$ is
not open in the Figs.~\ref{br} and \ref{br_tanb20}, because
$m_{\tilde{l}_L}^2 = m_{16}^2 + 0.8 M_2^2 -0.2 m_Z^2$ $\gsim
M^2_2>m^2_{\tchi^0_2}$. The decay into $\tilde{\tau}_1$ is also
suppressed due to the reduced left-right mixing. Thus, in the cMSSM
with $\mu\sim M$, the prospect of finding LFV at LHC is considerably
better compared with the MSUGRA model.

\begin{figure}[htbp]
\begin{center}
\includegraphics[width=6.0cm,angle=-90]{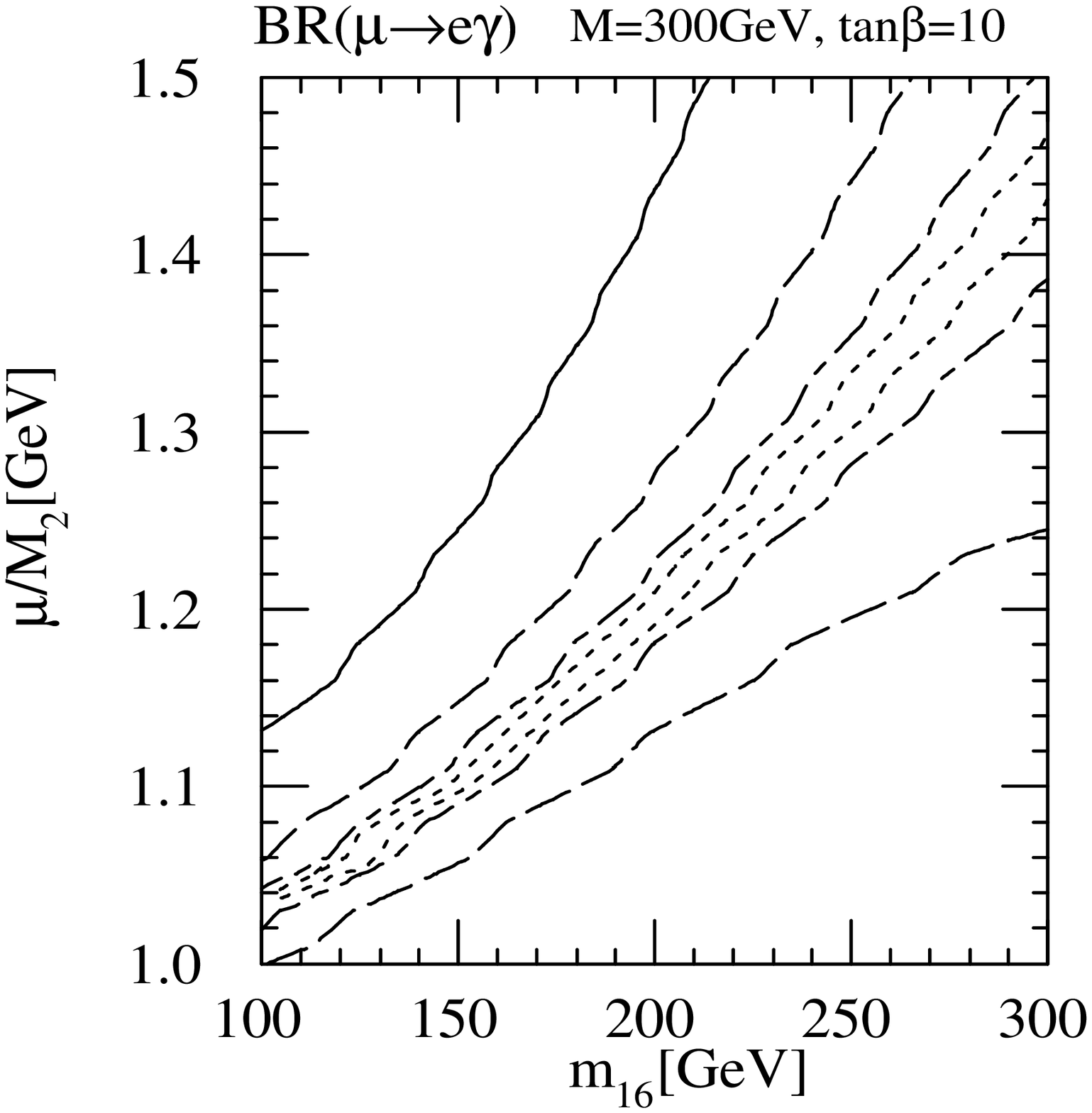}
\end{center}
\caption{\footnotesize 
Contours of constant 
$Br(\mu\rightarrow e \gamma)$ in $m_{16}$ and $\mu/M_2$ 
plane.
A solid line corresponds to $10^{-11}$, long-dashed lines $10^{-12}$ 
long and short dashed $10^{-13}$, short-dashed $10^{-14}$. 
Here, $M=300$GeV and $\tan\beta=10$.
\label{m0ovmulfv}}
\end{figure}

Varying the $\mu$ parameter also non-trivially changes the $\mu \to e
\gamma$ branching ratio.  In Fig.~\ref{m0ovmulfv}, we show the
contours of constant $Br(\mu\rightarrow e \gamma)$ in $m_{16}$ and
$\mu/M_2$ plane for the GUT scale boundary condition given in
Eq.~(\ref{soten}). Here we take $\tan\beta=10$ and $M=300$ GeV. We
assume the only non-zero mixing mass term between ${\tilde{e}_R}$ and
${\tilde{\mu}_R}$ exists at the GUT scale. We take the difference of
scalar masses at the GUT scale and the mixing as $\Delta m=1.2$ GeV
and $\sin 2\theta= 0.5$.

In the MSUGRA model, which corresponds to $\mu / M_2 \sim 1.5$,
$Br(\mu\rightarrow e \gamma)$ is less than $10^{-11}$ for $m_{16}=
m>210$ GeV and becomes minimum for $m_{16}\sim 300$ GeV.  The
suppression around $m_{16}\sim 300$ GeV is due to cancellation among
diagrams that will be discussed soon. Although LFV in the sfermion
decays does not suffer from such a cancellation among diagrams, the
decay of $\tilde{\chi}^0_2$ to $\tilde{l}$ is closed for $m_{16}=
m\gsim 210$ GeV. This shows that it is difficult to observe LFV at LHC
in the MSUGRA model.  The constraint from $Br( \mu \to e \gamma )$ is
strong to observe LFV at LHC in the MSUGRA model  for generic slepton
oscillation parameters as we see in the
next section.

The constraint weakens in the region where
$\tilde{\chi}^0_2\rightarrow \tilde{l}_R l$ is open, when $\mu$ is
smaller than the MSUGRA prediction. The parameter dependence of the
cancellation can be explained by the mass-insertion formula
\cite{Hisano:1996qq}, which is expressed by the off-diagonal component
$m^2_{\tilde{e}_R\tilde{\mu}_R}$ in the right-handed slepton mass
matrix. It occurs among the 4 different amplitudes where the chirality
flip occurs on either in the external or internal lines. Among them
two diagrams involving the left-right mixing or
lepton-slepton-Higgsino vertex have a common overall factor
$m^2_{\tilde{e}_R\tilde{\mu}_R}/m_{\tilde{e}_R}^4$ $\times
M_1\mu\tan\beta/m_{\tilde{e}_R}^2$ with opposite-sign
coefficients. When $\mu\tan\beta$ is large and the absolute values of
the two amplitudes are larger than the others, a nearly complete
cancellation could occurs when $\mu/m_{\tilde{l}_R}\sim \mu/m_{16}
\sim 1.5$, as in Fig.~\ref{m0ovmulfv}.  When $\mu<1.2 M_2$, the
cancellation occurs for $m_{16}<200$GeV where the decay of
$\tilde{\chi}^0_2$ to $\tilde{l}_R$ is kinematically open.

We have been discussing the case where LFV occurs due 
to the non-zero  $m^2_{\tilde{e}_R\tilde{\mu}_R}$. We will comment on 
the other cases in the last section. 

\section{Potential of the LFV search at LHC in the cMSSM model}

In the limit of lepton-flavor conservation, the process
$\tilde{\chi}_2^0\rightarrow \tilde{l}l\rightarrow ll
\tilde{\chi}_0^0$ at LHC has been studied by many authors
\cite{TDR}. The distribution of the lepton-pair invariant mass
($m_{ll}$) in the final state is given by
\begin{eqnarray}
\frac{d \Gamma_{l^- l^+}}{d m_{ll}}
&\propto&
\left\{
\begin{array}{ll}
m_{ll}&(0 \le m_{ll} \le m_{ll}^{\rm max})\\
0      &(m_{ll}^{\rm max} < m_{ll})
\end{array}
\right.
\label{mlldis_intro}
\ ,
\end{eqnarray}
where the edge $m_{ll}^{\rm max}$ is expressed by the
slepton mass $m_{\tilde{l}}$ and the neutralino masses
$m_{\tilde{\chi}_{1,2}^0}$ as follows:
\begin{eqnarray}
(m_{ll}^{\rm max})^2= {m}^2_{\tilde{\chi}^0_2}
(1-\frac{{m}^2_{\tilde{l}}}{\mzii^2})
(1-\frac{\mlsp^2}{{m}^2_{\tilde{l}}})
\ . 
\end{eqnarray}
This decay process would be identified through the edge in
Eq.~(\ref{mlldis_intro}).  The main background comes from uncorrelated
leptons from different squark or gluino decay chains. Fortunately, the
background are estimated using the $e^\pm\mu^\mp$ distribution, and
the distribution is smooth and decreases monotonically as $m_{e\mu}$
increases.  Thus, the background can be subtracted, and the
subtracted distribution has a canonical structure as
$d\Gamma/dm_{ll}\propto m_{ll}$ and terminates at $m_{ll}^{\rm max}$.

The signature of LFV on the process is the
edge structure of $e^{\pm}\mu^{\mp}$ distribution on top of the
accidental leptons from uncorrelated sources, since such an efficient
subtraction method as above does not exist. The level of
the signal and the background would be estimated if the production cross
sections, acceptance of the signal and the background, and the background
distribution could be estimated. This is of course possible by doing the MC
simulation for each model parameters, however, we present
semi-analytical approach in this section. In the following
subsections, we will discuss the production cross section of the SUSY
particles, level of the signal and the background, and the background
distribution, and show the experimental reach of the LFV search at LHC 
at the end.
 
\subsection{Production cross section of the SUSY particles}

We start our discussion from an estimation of the squark and gluino
production cross sections.  The second-lightest neutralino
$\tilde{\chi}_2^0$ is produced through the cascade decays of
$\tilde{q}_L$ or $\tilde{g}$.  The signal rate of the LFV decay
$\tilde{\chi}^0_2\rightarrow l'l\tilde{\chi}^0_1$ depends on the
production cross section significantly, because it reduces very
quickly with increase of the squark and gluino masses.

We use the ISAJET version 7.51 \cite{Baer:1999sp} to estimate the
production cross sections.\footnote{The parton distribution is given
by CTEQ3L.} We are interested in the region of parameter space where
$m\lesssim M$ so that the $\tilde\chi^0_2$ decay to $\tilde l$ is
open. In this range, we choose the four parameter sets A)--D) in
Table~1 as samples and derive a fitting functions for the production
cross sections in Table~2.\footnote{
We omit the cross sections of the stop- and sbottom-pair production 
since they give negligible contributions to the $\mu e$ background.
}
\begin{table}[t]
\begin{center}
\begin{tabular}{|c||c|c|c|c|c|}
\hline
& $M$(GeV) & $m$ & $\mgl$& $\msql$ & $\msti$   
\cr
\hline
A)& 300 & 100 & 706 & 633 & 470 
\cr
B)& 350 & 125 & 818 & 729 & 549 
\cr
C)& 400 & 150 & 915 & 824 & 627 
\cr
D)& 500 & 175 & 1135 & 1012 & 781 
\cr
\hline 
\end{tabular}
\end{center}
\caption{Description of the set of parameters A)-D). We chose
$\tan\beta=10$ and $A=0$ in the MSUGRA model. The GUT scale gaugino
mass $M$, the scalar mass $m$, the squark and gluino masses are given.}
\end{table}
\begin{table}[t]
\begin{center}
\begin{tabular}{|c||c|c|c|c|c|}
\hline
(pb)&$\sigma$(total)& $\sigma(\gl\gl)$& 
$\sigma(\gl\sq)$+$\sigma(\gl\sq^\star)$&
$\sigma(\sq\sq)$& $\sigma(\sq\sqb)$
\cr
\hline
A)& 25.0   &3.03& 13.04 & 4.70 & 2.89
\cr
B)& 11.2   & 1.11& 5.67 & 2.41 & 1.41 
\cr
C)& 5.74  & 0.45 & 2.87 & 1.44& 0.67 
\cr
D)& 1.60  & 0.09 & 0.76 & 0.51 & 0.19
\cr
\hline
\end{tabular}
\end{center}
\caption{Production cross sections for the parameters A)-D). }
\end{table}
The fitting functions will be used later to estimate
the number of the signal and background $e^{\pm}\mu^{\mp}$
events. 

Since the squark and gluino masses are quite close to each
other for those points, we fit the production cross sections by the
following simple functions of $m_{\tilde{g}}$,
\begin{eqnarray}
\sigma ({\tilde{q}\  {\rm or}\   \tilde{g}}) 
&=& a_1 (\mgl/{\rm TeV})^{-a_2} \ \ ({\rm pb}).
\label{fit}
\end{eqnarray}
The production cross sections are fitted very nicely by the parameters
$a_1$ and $a_2$, listed in the Table~3. 
When the scalar masses are changed in
a relevant region of the parameter space, the cross sections are changed
within only 20\%, which is within QCD uncertainties. 
\begin{table}
\begin{center}
\begin{tabular}{|c||c|c|c|c|c|}
\hline
&$\sigma({\rm total})$& $\sigma(\gl\gl)$& 
$\sigma(\gl\sq)$+$\sigma(\gl\sq^\star)$ &
$\sigma(\sq\sq)$& $\sigma(\sq\sqb)$
\cr
\hline
$a_1$ & 3.34& 0.23&  1.74&  0.92&  0.39\cr
$a_2$ &5.78& 7.407& 5.780& 4.677& 5.733\cr
\hline
\end{tabular}
\caption{Results of the fit of the production cross sections 
to the functions given in Eq.~(\ref{fit}). }
\end{center}
\end{table}

Note that the scaling parameter $a_2$ depends on the production modes
substantially. Processes involving initial state gluon(s) such as
$\sigma(\tilde{g}\tilde{q})$, $\sigma(\tilde{q}^\star\tilde{g})$ reduces
quickly compared with $\sigma(\tilde{q}\tilde{q})$ production when
gluino and squark masses become heavier. This comes from a fact that
the parton distributions of the gluon and sea quarks are generally
softer than valence quarks.

The $\tilde{u}$/$\tilde{d}$ and $\tilde{q}_L/\tilde{q}_R$ ratios
affect the level of $\mu e $ background. The $\tilde{u}_L\tilde{d}_L$,
$\tilde{u}_L\tilde{u}_L^\star$, and $\tilde{d}_L\tilde{d}_L^\star$
production are likely to contribute to the signal compared with the
other squark productions, because the up-type squark $\tilde{u}_L$
would decay into $\tilde{\chi}^+_1$ producing $l^+$ while
$\tilde{d}_L$ producing $l^-$, and $\tilde{q}_R$ decay mostly into
$\tilde{\chi}_1^0$ when $\tilde{\chi}^+_1$, $\tilde{\chi}^0_1$,
$\tilde{\chi}^0_2$ are gaugino-like. Also, note that the production is
dominated by valence $u$ and $d$ components of parton distribution
function as in Table~2. Thus, we adopt a simple approximation that
$\tilde{u}:\tilde{d}= 2:1$, $\tilde{u}^\star:\tilde{d}^\star= 1:1$, and
$\tilde{q}_L:\tilde{q}_R= 1:1$ in any squark and/or anti-squark
production cross sections.\footnote{
In the MC simulation, the ratio of production cross sections,
$\sigma(\tilde{u}\tilde{u}):$ $\sigma(\tilde{u}\tilde{d}):$
$\sigma(\tilde{d}\tilde{d})$, is 4.4:3.9:1 for point A) and and
7:5.3:1 for point D).  For $\tilde{q}\tilde{q}^\star$ production cross
sections, the ratio $\sigma(\tilde{u}\tilde{u}^\star):$
$\sigma(\tilde{u}\tilde{d}^\star):$ $\sigma(\tilde{d}\tilde{u}^\star):$
$\sigma(\tilde{d}\tilde{d}^\star)$ is 5.7:4.2:1:3.9 for point A). And,
$\sigma(\tilde{q}_L\tilde{q}_L): \sigma(\tilde{q}_L\tilde{q}_R):$
$\sigma(\tilde{q}_R\tilde{q}_R)$=1:1.4:1.2 and
$\sigma(\tilde{q}_L\tilde{q}^\star_L) : \sigma(\tilde{q}_L\tilde{q}^\star_R):
\sigma(\tilde{q}^\star_L \tilde{q}_L): \sigma(\tilde{q}_R\tilde{q}_R)$=
1:0.93:0.90:1.15 for point A).}

\subsection{Level of the signal and the background and the acceptance} 

Next, we will estimate the level of the signal and the background, and
evaluate the acceptance.  
For this purpose, we use the MC data for the following parameters\footnote{
The mass parameters and relevant branching ratios are given in the
Appendix B, which are calculated by the ISAJET.
};
\begin{description}
\item[point I)] $\mu=497.87$GeV, $\tan\beta=2.1$, $M=300$ GeV,
and $m=100$ GeV in the MSUGRA \cite{Nojiri:2000wq},
\item[point II)] $\mu=199.85$GeV $\tan\beta=10$, $M=250$ GeV, and $m_{16}=90$ 
GeV in the cMSSM \cite{Drees:2001he}. 
\end{description}

  To estimate level of the background, we have to calculate branching
ratios $Br(\tilde{g} \rightarrow l^{\pm}X)$, $Br(\tilde{g} \rightarrow
l^+l^{'-}X)$, $Br(\tilde{q}\rightarrow l^{\pm}X)$, and
$Br(\tilde{q}\rightarrow l^+l^{'-}X)$. We find that 
$\tilde{g}$ has a large branching ratio into multiple leptons. For
example, $\tilde{g}\rightarrow\tilde{t}t$ and
$\tilde{g}\rightarrow\tilde{b}_1b$ followed by $\tilde{b}_1\rightarrow
t \chi^+$ or $\tilde{b}_1\rightarrow \tilde{t} W^- $ have fractions
15\%, 6\%, and 4\%, respectively, for point I). This results in a
high probability to get multiple leptons in the final states;
$Br(\tilde{g}\rightarrow (\tilde{t},\tilde{b}_1, \tilde{b}_2)
\rightarrow l)= 4$\% and $Br(\tilde{g}\rightarrow
(\tilde{t},\tilde{b}_1, \tilde{b}_2) \rightarrow l'l\ or\
l\tau(\rightarrow l'))= 2$\%. Thus, the $\tilde{g}$ production may be
a significant source for the background for the LFV search compared
with the squark-pair production processes, since the squark-pair
production processes are required to have two cascade decays involving
a lepton each so that they contribute to the background.
 
The chargino production is also a significant source of the
background. A chargino may decay into $W$, $\tilde{\nu}$, and
$\tilde{\tau}_1$, followed by decay into leptons in the final
state.\footnote{
For example, 89\% of chargino decays into $\tilde{\tau}_1$ for
point I).
} 
A produced tau lepton may further decays into $e$ or $\mu$, whose
branching fraction is about 35\%.  The second-lightest neutralino
$\tilde{\chi}^0_2$ may also decay into $\tau\tilde{\tau}_1$. While the
momentum of $e$ or $\mu$ from the tau-lepton decay tends to be low,
the acceptance of such leptons would not be negligible for the adopted
lepton $p_T$ cut as low as 10 GeV. For the case where $\mu\sim M_2$,
one should also take care of the decay of $\tilde{q}$ into
$\tilde{\chi}^0_{3(4)},\tilde{\chi}^{\pm}_2$. They are also calculated
and added in our background estimation.

The total $e^{\pm}\mu^{\mp}X$ and $l^+l^-X $ events before cuts are
shown in Table~4. The numbers will be compared with the MC results in
order to derive the acceptance. They are obtained by calculating the
gluino and squark branching ratios into the lepton(s) using the major
branching modes mentioned above, and multiplying our fitted production
cross sections described previously to the branching ratios.  This
semi-analytical calculation is checked by independent toy MC
simulations which includes all decay steps. We include the
contribution from $\tilde{q}\tilde{q}$, $\tilde{q}\tilde{q}^\star$ and
$\tilde{q}\tilde{g}$ production.  $N_{l l'}$ is the number of the
$e^{\pm}\mu^{\mp}$ events coming from primary leptons from $W$ or
$\tilde{l}$ decay.  $N_{l\tau}$ is the number of the events with one
primary leptons and one lepton from $\tau\rightarrow e$ or
$\mu$. $N_{\tau\tau}$ is the number of $e^{\pm} \mu^{\mp}$ events from
the leptonic decay of two $\tau$'s; here we omit the contribution from
$\tchi^0_i\rightarrow\tau^+\tau^-\tchi^0_1$, because the $m_{l'l}$
distribution are substantially softer than expected LFV signal
$\tchi^0_2\rightarrow \tilde{l}'' l \rightarrow l'l\tilde{\chi}^0_1$.
Note that our calculation does not include the probability that $3l$,
or $4l$ events are accepted as two lepton events etc.

Point II) is substantially lepton rich compared with point I)
due to the enhanced branching ratio into sleptons. This is typical for
parameters with $\mu\sim M_2$ with light sleptons as shown 
in Fig.~\ref{br}.

Now we can estimate the acceptance of events involving $l^+l^-$,
$e^{+}\mu^-$, or $e^{-}\mu^+$. Here we use MC data produced and
simulated by the ISAJET and the ATLFAST. $2\times10^6$($1\times10^7$)
events are generated for point I) (point II)) corresponding to
95(196)fb$^{-1}$ of the integrated luminosity. In the simulation, we
adopted the cuts given in \cite{Bachacou:2000zb}
\begin{itemize}
\item $E^T_{\rm miss}>$ max $(100 {\rm GeV}, 0.2 M_{\rm eff})$ , 
\item $P^T_{j1}>100$ GeV, $P^T_{j_2,j_3,j_4}>50$GeV,
\item $M_{\rm eff }>400$ GeV,
\item a pair of isolated opposite sign leptons with $P^T_l>10$ GeV. 
\end{itemize}
Those cuts are chosen to reduce backgrounds from QCD processes to a
negligible level. After the cuts, the $e^\pm\mu^\mp$
events are reduced to 3609 events for point I), and 42721 events
for point II).  (See $N(e\mu)$(MC) in Table 4.) By comparing
$N(l'l)$, $N(l\tau)$, $N(\tau\tau)$ with $N(e\mu)$(MC), we estimate
the acceptance of uncorrelated $e\mu$ events at the ATLAS detector;
19\% for  point I), and 22\% for point II).

\begin{table}
\begin{center}
\begin{tabular}{|c||c|c|c|c|}
\hline
point &   $N(e\mu)$ (MC) & $N(l'l)$&  $N(l\tau)$ & $N(\tau\tau)$ 
\\ \hline
point I)  &  3609 & $1.49\times 10^4$  &
 $3.61\times 10^3$&
$3.2 \times 10^2$ \\
%mixed   & 42721(0 to 200 GeV  & 11460 & 94144 & 49817\\
point II)  & 42721  &$ 3.1 \times 10^4$  & $10.8\times 10^4$ &
$5.2\times 10^4$\\
\hline
point I) $b$ veto & &$1.58\times 10^3$ & $5.4\times 10^2$& $0.67\times 10^2$ \\
point II) $b$ veto  & &$3.78\times 10^3$ & $3.85\times 10^4$ & $2.20\times
10^4$ 
\\\hline
\end{tabular}
\end{center}
\caption{\footnotesize 
Number of events with $e^{\pm}\mu^{\mp}$ before cuts from
different decay processes $N(l'l)$, $N(l\tau)$ and $N(\tau\tau)$ when
lepton flavor is conserved. (See text.)  The numbers are for
lepton-flavor conservation. $N(e\mu)$(MC) is the numbers of accepted
events with $e^{\pm}\mu^{\mp}$ satisfying $0$GeV$<m_{e\mu}<200$GeV,
obtained by the ISAJET+ATLFAST simulation. }
\end{table}

The acceptance of LFV signal may be also estimated by looking at the
the acceptance of opposite-sign same-flavor leptons in the same
simulation. Note that the LFV signal from the $\tchi^0_2$ decay has exactly
the same kinematics to the lepton-flavor-conserving decay. We list in
Table~5 the result of the MC simulation of the same samples as that of
Table~4 and the estimated $l^+l^-$ production from the $\tchi^0_2$ decay
before the cut.  The acceptance of the process is 28\%. This is higher
than that of $e^{\pm}\mu^{\mp}$ events under the same cut. To be
conservative, we adopt constant 25\% acceptance for both the signal and
the background.

To suppress the background furthermore, the $b$-jet veto may be efficient
while we will not use it in this article. In the Tables~4 and 5, we
show numbers of events which do not involve the third-generation
squarks in the cascade decays.  The events involving $\tilde{t}$ or
$\tilde{b}$ could be removed by the $b$-jet veto. If the efficiency of the
veto is ideal, $N(l'l)$ is reduced by an order as in Table~4, while the number
of the correlated leptons from $\tchi^0_2\rightarrow \tilde{l}$ decays
does not change in Table~5 by the $b$-jet veto.  However, the
rejection factor depends on the decay patterns. For example,
$\tilde{\chi}^+_1\rightarrow\tilde{\tau}$ dominates the chargino decay
for point II), and the decay $\tilde{q}_L\rightarrow
\chi^+_1$ becomes the efficient source of $\tau$ lepton. This means
that $N(l\tau)$ is not reduced so much by the $b$-jet vet in point II).  

\begin{table}
\begin{center}
\begin{tabular}{|c| c| c|}
\hline
 point & $l^+l^-$ events before cut & accepted $l^+l^-$ by MC \cr
\hline
point I) (95fb$^{-1}$)& $8.72\times 10^4$  &
$2.55\times 10^4$ \\
point II) (196 $fb^{-1}$)&   $8.97\times10^5$ 
&$2.55\times 10^5$ \\
\hline
point I) $b$ veto& $6.51\times 10^4$ & \\
point II) $b$ veto  & $6.96\times 10^5$ & \\
\hline
\end{tabular}
\end{center}
\caption{\footnotesize 
The  number of $\tchi^0_2\rightarrow l^+l^-\tchi^0_1$  when 
LFV is absent. The accepted $l^+l^-$ events in  MC simulations 
are also given in the table.}
\end{table}

%\noindent
\subsection{Background distribution}

The signal distribution from $\tchi^0_2$ decay increases with
$m_{l'l}$ as in Eq.~(\ref{mlldis_intro}) and the distribution has an edge
determined by the neutralino and the sfermion masses. On the other
hand, backgrounds come from $t$, $W$, and  $\tilde{\chi}^{\pm}_i$ decays
and do not have the edge structure. They reduce rather quickly as $m_{l'l}$
increases. Therefore it is better to use the data near $m^{\rm
max}_{ll}$ so that $S/N$ ratio maximizes. We should note that the
position of the $m_{l'l}$ edge is known precisely from the same-flavor
opposite-sign $m_{ll}$ distribution.

\begin{figure}[htbp]
\begin{center}
\includegraphics[width=7.5cm,angle=-90]{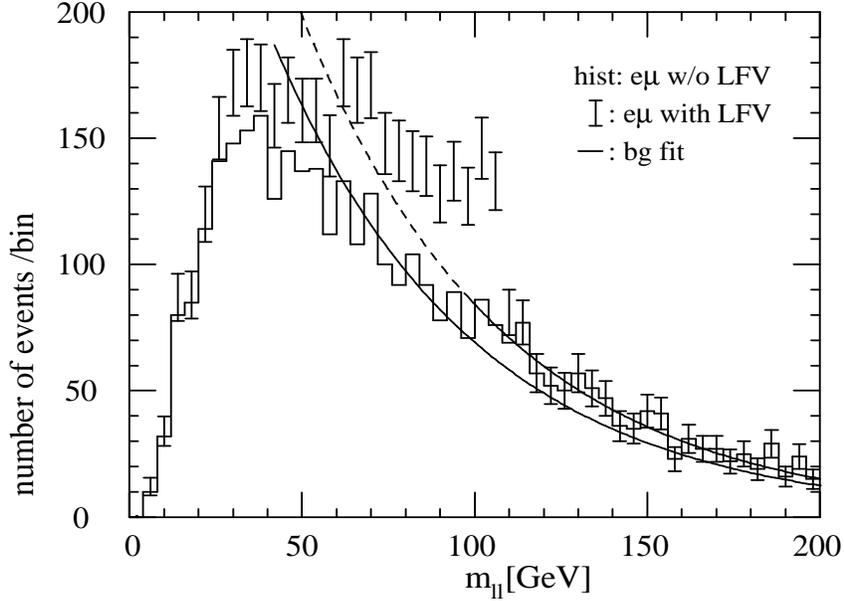}
\end{center}  
\caption{\footnotesize The $m_{e\mu}$ distribution for point I).  The
data corresponds to integrated luminosity of 95 fb$^{-1}$ and standard
cuts are applied (see text). The histogram shows the distribution
without LFV, while bars are number of events and the error with
$\tilde{\mu}$-$\tilde{e}$ mixing. In the plot, 1/30 of
$\tilde{\chi}^0_2\rightarrow \tilde{l}''l$, $\tilde{l}''\rightarrow
\chi^0_1 l'$ decay chain is assumed to go to the e$\mu$ channel.  Two curves
are fits to the background distribution in the region $m_{ll}=40$--$200$
GeV(solid) and $m_{ll}=100$--$200$(dashed then solid). We use $c=12.1$. 
\label{bgfit1}}
\end{figure}

\begin{figure}
\begin{center}
\includegraphics[width=7.5cm,angle=-90]{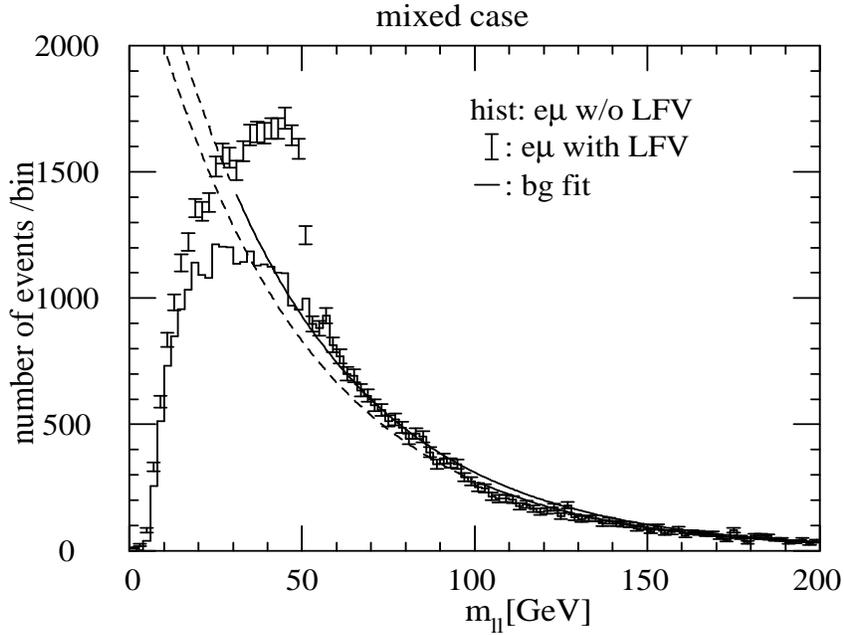}
\end{center}
\caption{Same as Fig.~\ref{bgfit1}, but for point II).
The integrated luminosity is $196$ fb$^{-1}$, and $c=13.7$. 
\label{bgfit_mixed1}}
\end{figure}

To estimate the number of the background events near the edge, we again 
use the MC data and fit it to the following fitting 
function,
\begin{equation} 
\frac{d\Gamma}{dm_{ll}} =k(m_{\tilde{g}}) \exp\left( -
\frac{c}{m_{\tilde{g}}} \times m_{ll}
\right). \label{mllfit}
\end{equation} 
The background distributions and the fitting curves are shown in
Figs.~\ref{bgfit1} and \ref{bgfit_mixed1}. 
The data between 40 to 200 GeV and 100 to 200 GeV are
used for the fit. The best fit is obtained $c=10.4$ for point I) and
$c=13.7$ for point II) for $30$ GeV $<m_{e\mu}<200$ GeV.
The fitting function reproduces the large $m_{l'l}$ region reasonably
well, but it fails significantly in the small $m_{l'l}$ region.
It is natural that the distribution has a certain peak, 
which must be proportional to a typical momentum of leptons of the
uncorrelated production process, such as the half of $W$ boson mass.
The distribution beyond this peak must be more sensitive to typical
momentums of $W$, $t$, or $\tchi^+_1$, which may depend on the gluino
and squark masses. This is the reason we choose Eq.~(\ref{mllfit}) as a 
fitting function. The average value for $c$ of those two points, $c=
12.1$, is used for our background estimation. For the plot, $k(M)$ may
be fixed so that overall normalization agrees for the region used for
the fit.


\subsection{Significance of LFV at LHC}

Having gone through all estimation needed, we now calculate
significance of the LFV signal at LHC.  Numbers of the signal
($N^{sig}$) and background ($N^{bg}$ ) are estimated by (the fitted
cross sections) $\times$ (the branching ratios) $\times$ (the overall
acceptance 25\%) $\times$ (the integrated luminosity).  We assume the signal
distribution in Eq.~(\ref{mlldis_intro}) and background distribution
in Eq.~(\ref{mllfit}). We determine the overall normalization factor
of Eq.~(\ref{mllfit}) so that number of background above $m_{l'l}>20$
GeV agrees with the estimation.

We define the 
$\Delta \chi^2$ using the estimated signal and background events between 
\begin{equation}
{\rm max}\left(30 {\rm GeV}, \frac{2}{3} 
m_{ll}^{\rm max}\right)<m_{ll} < m_{ll}^{\rm max},
\label{mllsignal}\end{equation}
and calculates 
\begin{equation}           
\Delta\chi^2= \sum_i  \frac{(N^{\rm sig}_i)^2}{N^{\rm sig}_i+ N^{\rm bg}_i}
\label{chi2}
\end{equation}
for the bin size $2n$ GeV, where integer  $n$  is determined 
so that ${\rm max}(N^{\rm sig}_i)>10 $.
Eq.~(\ref{chi2})
expresses the statistical significance of the signal after the
subtraction of expected background.\footnote{
We assume no error for the background shape. 
}
In the experimental situation, one may
determine the background distribution from the real data, when the
events above the $m_{ll}^{\rm max}$ may be used to make a simple
extrapolation as suggested in Eq.~(\ref{mllfit}).\footnote{
Alternatively, one can constrain the MSSM parameters as
model-independent as possible, so that the background distribution can
be determined model-independently. Note that the nature of the
third-generation sfermions is important since the substantial fraction
of the $e^{\pm}\mu^{\mp}$ events might come from $\tilde{g}\rightarrow
\tilde{t}_1t$ or $\tilde{b}b$ followed by their cascade decays to $W$
etc.  The sbottom mass may be reconstructed from the distribution of
the events with $b$ jet \cite{TDR}.  Attempts to reconstruct stop
decays may be found in \cite{TDR, kawagoe}
}

\begin{figure}[htb]
\begin{center}
\includegraphics[width=7.5cm,angle=-90]{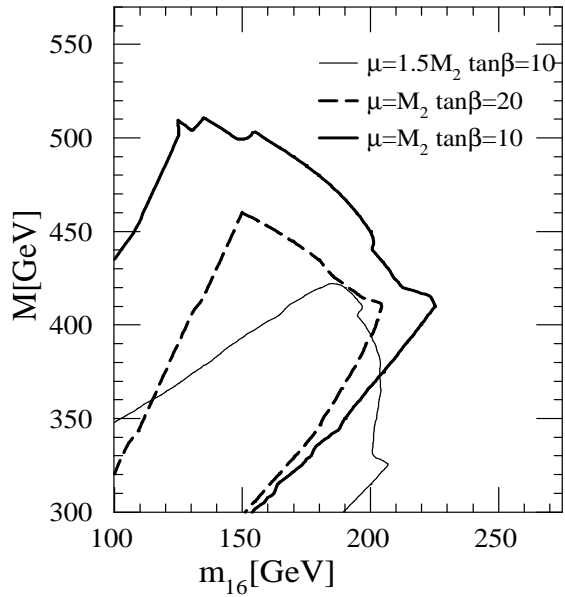}
\end{center}  
\caption
{ \footnotesize $\sqrt{\Delta \chi^2}=5$ contours for the LFV
discovery.  The thick solid line is for $\mu=1.5 M_2$ and
$\tan\beta=10$ in the cMSSM, the thick dashed line for $\mu=M_2$ and
$\tan\beta=20$, and the solid line for $\mu=M_2$ and
$\tan\beta=10$. We fix the $\ser$-$\smur$ mixing angle $\theta$ as
$\sin 2\theta=0.5$ and the slepton mass difference $\Delta m=1.2$ GeV
at the GUT scale.
\label{chi2plot}}
\end{figure}

In Fig.~\ref{chi2plot}, we show the contours of constant
$\Delta\chi^2$ in $m_{16}$ and $M$ plane. $\Delta\chi^2=25$ $\sim$ $5
\sigma$ contours correspond to 70 signal $e^{\pm}\mu^{\mp}$ events in
the signal region in Eq.~(\ref{mllsignal}). Here, the integrated
luminosity is 100fb$^{-1}$. The SUSY background is roughly of the same
order as the signal.  In Ref.~\cite{PSUSY}, 120 total SM background
events are expected for 30 fb$^{-1}$ under the cuts 1)
$E^T_{miss}>300$ GeV, 2) $p^T_{l} > 10$ GeV, and 3) two jets with
$p^T_{jet}>150$ GeV.  The level of the background in the signal region
is of the same order as that of the SUSY background. The significance
beyond this $5\sigma$ contour is therefore the subject of more careful
MC simulations both for SUSY and SM backgrounds.

When $\mu\sim M_2$, the parameter space covered by LHC extends,
compared with the MSUGRA model ($\mu\sim 1.5 M_2$), due to the
enhanced branching rates of $\tilde{\chi}^0_2$ to $\tilde{l}_R$ as we
see in the previous section.  We can see another qualitative
difference in $m_{16}\ll M$ region. For the MSUGRA case, the small
$m_{16}$ region cannot be reached because $\tchi^0_2\rightarrow
\tilde{l}_L$ dominates. The search region is extended to this region
for the cMSSM with $\mu\sim M$ because $m_{\tilde{\chi}^0_2}<M_2$ and
$\tilde{\chi}^0_2$ could not decay into $\tilde{l}_L$ for $\mu\sim M$.

\begin{figure}[p]
\begin{center}
\hspace*{1.5cm}
\includegraphics[width=11cm]{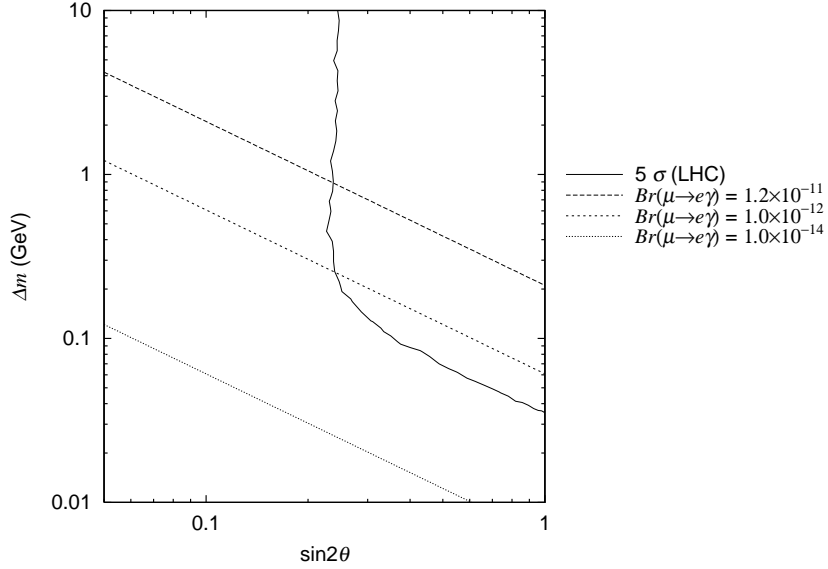}
\end{center}  
\caption
{\footnotesize The LHC reach and the line of the constant $Br(\mu \to
e \gamma)$ in the MSUGRA model are shown.  Here, $\tan \beta = 10$,
$A=0$, $m = 100$ GeV, and $M=300$ GeV.
\label{osc_sugra}}
\end{figure}

\begin{figure}[hp]
\begin{center}
\hspace*{1.5cm}
\includegraphics[width=11cm]{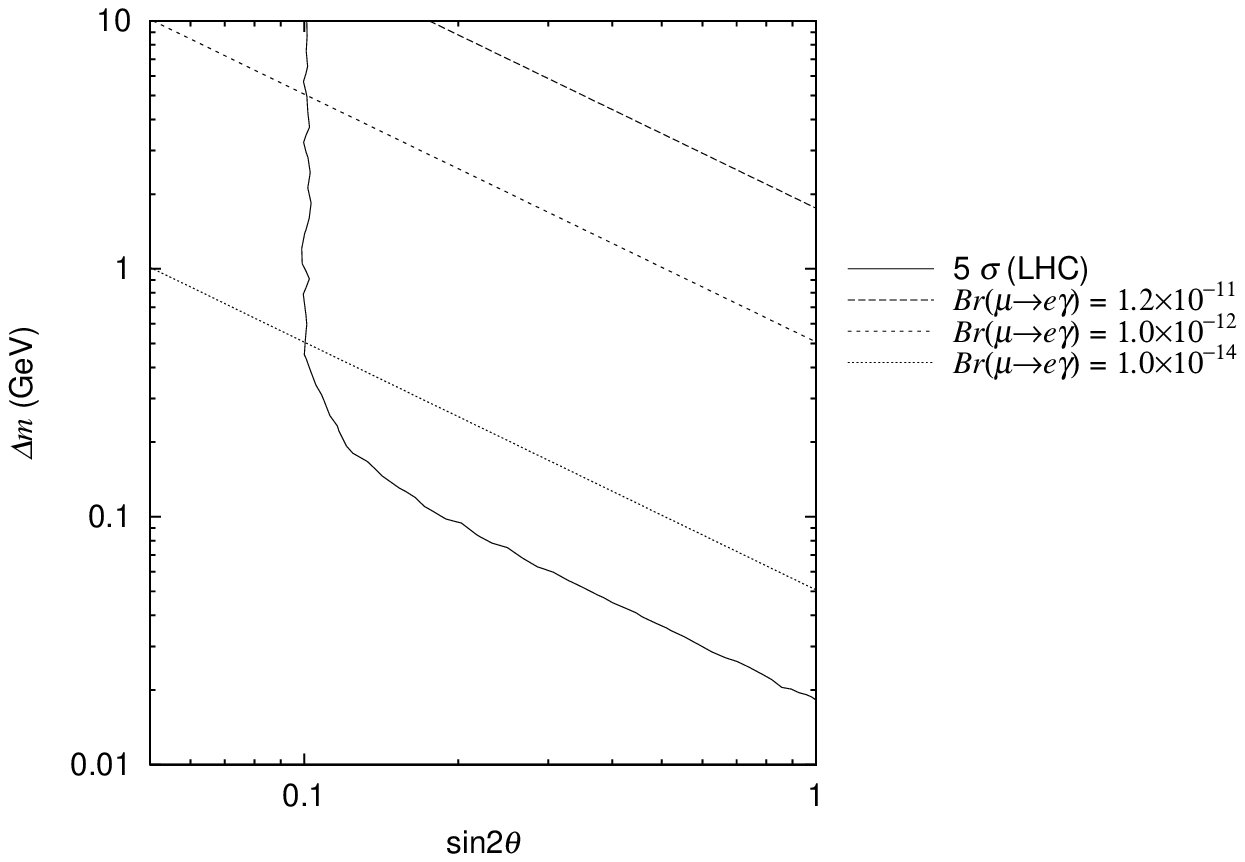}
\end{center}  
\caption
{\footnotesize
The LHC reach and the line of the constant $Br(\mu \to e \gamma)$
in the cMSSM are shown.
Here, $\mu = M_2$, $\tan \beta = 10$, $A=0$, $m_{16} = 100$ GeV,
and $M=300$ GeV.
\label{osc_mix}}
\end{figure}

We also estimate the LHC reach for generic oscillation parameters.  In
Fig.~\ref{osc_sugra}, we plot the $5\sigma$ contour and the line of
$Br(\mu \to e \gamma)$=$1.2 \times 10^{-11}$, $1.0 \times 10^{-12}$,
and $1.0 \times 10^{-14}$ in the parameter space of $\sin 2 \theta$
and $\Delta m$ at the GUT scale.  We use the the MSUGRA model with
$\tan \beta = 10$, $A=0$, $m = 100$ GeV, and $M=300$ GeV.  We can
see that the most part of the parameter range where LFV can be
observed at LHC is already excluded by the $Br(\mu \to e \gamma)$
constraint, and the remained region will be covered by next-generation
experiments.  The situation changes for the cMSSM case with $\mu=M_2$.
The corresponding figure is shown in Fig.~\ref{osc_mix}.  Because of
the change of the decay kinematics, the wider range of the parameter
space is covered by the LHC than that in the MSUGRA case.  $Br(\mu \to
e \gamma)$, in contrast, becomes small due to the cancellation between
the diagrams.  It follows that LHC might be more advantageous to
observe LFV than the $\mu \to e \gamma$ decay search, especially in
the cMSSM.


\section{Conclusions and Discussion}

In this paper we investigate the potential of LHC to find LFV in
$\tchi^0_2\rightarrow\tilde{l}''l$, $\tilde{l}''\rightarrow l'
\tchi^0_1$. Here we studied it in a general model where the $\mu$
parameter is independent of the gaugino mass $M$ by allowing the
non-universal GUT scale Higgs masses. An approximated universality of
squark and slepton masses is imposed. We find LHC would be able to
find the LFV mixing between the first and second generation in the
right-handed slepton masses, $m^2_{\tilde{e}_R\tilde{\mu}_R}$ while
the $Br(\mu\rightarrow e \gamma)$ is undetectably small.

\begin{figure}
\begin{center}
\includegraphics[width=7.5cm,angle=270]{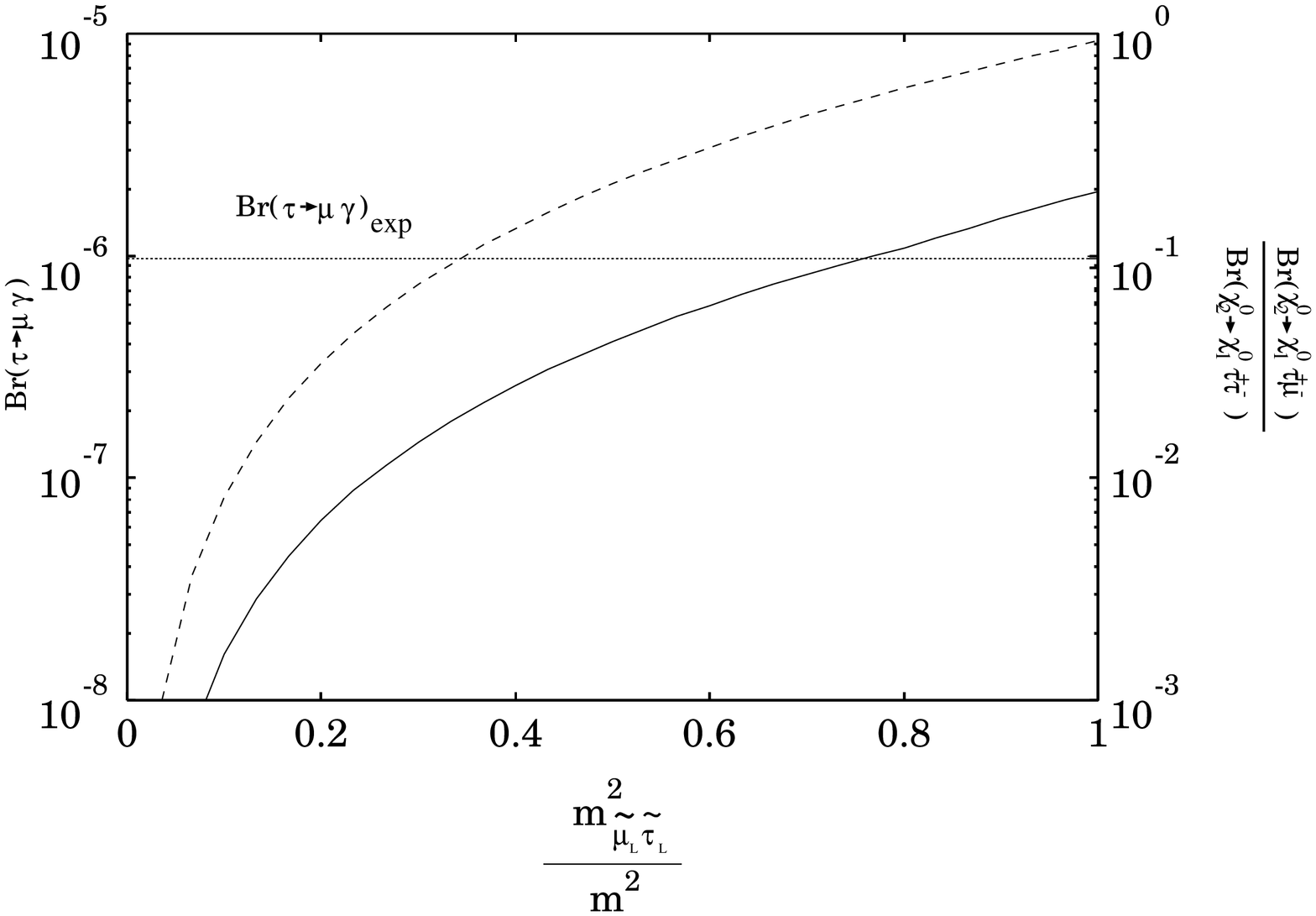}
\end{center}
\caption{ 
\footnotesize
$Br(\tchi^0_2\rightarrow\tchi^0_1 \tau\mu)$ normalized by
$Br(\tchi^0_2\rightarrow\tchi^0_1 \tau\tau)$ (the solid line)
and $Br(\tau\rightarrow \mu\gamma)$ (the dashed line)
as functions of $m^2_{\tilde{\mu}_L\tilde{\tau}_L}/m^2$.
Here, we take $m=130$GeV, $M=250$GeV, $A_0=300$GeV and $\tan\beta=10$.
\label{taumu}}
\end{figure}

LFV in left-handed slepton mass matrix as
$m^2_{\tilde{\mu}_L\tilde{\tau}_L}$ might be more motivated when the
data from atmospheric neutrino is considered
\cite{Hisano:1998wn}\cite{Hisano:1998fj}. We note that the
cancellation among the LFV diagrams is unlikely when only the
left-handed slepton mass is the source of LFV. We show the relation
between $Br(\tau\rightarrow \mu\gamma)$ (the dashed line) and
$Br(\tchi^0_2\rightarrow\tchi^0_1 \tau\mu)$ normalized by
$Br(\tchi^0_2\rightarrow\tchi^0_1 \tau\tau)$ (the solid line) in
Fig.~\ref{taumu} when $m^2_{\tilde{\mu}_L\tilde{\tau}_L}$ is the
unique source of LFV.  Hinchliffe and Paige stated that it is possible
to observe LFV for $Br(\tilde{\chi}^0_2 \to \tilde{\chi}^0_1 \tau \mu)
\gsim 0.01$ in this figure.  We can see in Fig.~\ref{taumu} that the
reach of LHC corresponds to $Br(\tau \to \mu \gamma) \simeq 10^{-6}$,
which is also within the range of the $\tau \to \mu \gamma$ search at
the KEKB experiment \cite{ohshima}.

Finally when several off-diagonal scalar masses
$m^2_{\tilde{l}'\tilde{l}}$ are non-zero, $Br(l\rightarrow l'\gamma)$
could show the very complicated structure, therefore negative results
in the rare decay search at low energy do not necessary constrain the
processes involving $\tilde{l}$ decays at the high energy
colliders. Especially, the LFV processes, $\mu\rightarrow e \gamma$
and $\mu N\rightarrow e N$, may be more sensitive to all LFV slepton
masses, compared with those involving the third-generation sleptons,
such as $\tau\rightarrow\mu\gamma$.  The size of
$m^2_{\tilde{e}\tilde{\tau}}$ and $m^2_{\tilde{\mu}\tilde{\tau}}$ is
less constrained compared with $m^2_{\tilde{e}\tilde{\mu}}$, and then
the LFV through $\tilde{\tau}$ could overcome direct
$\tilde{\mu}$-$\tilde{e}$ mixing.  For example, even when only the
left-handed sleptons have the LFV masses, there is a cancellation in
$\mu\rightarrow e \gamma$ among the diagrams in some specific
parameter space \cite{Ellis:2001xt}.

\section*{Acknowledgments}
We would like to thank Dr. D.~Toya and Dr. K.~Kawagoe, who provided
simulation data important for our study. This work is supported in
part by the Grant-in-Aid for Science Research, Ministry of Education,
Science and Culture, Japan (No.~12047217 for M.M.N and Priority Area
707 `Supersymmetry and Unified Theory of Elementary Particles'
No.~13001292 for J.H.) and JSPS Research Fellowships for Young
Scientists (R.K.).

\section*{Appendix A:Formula for LFV processes} 
\setcounter{equation}{0}
\renewcommand{\theequation}{A.\arabic{equation}}

First, we present our formula for the LFV decay of the second-lightest
neutralino, $\tilde{\chi}_2^0\rightarrow l^{'-}
(\tilde{l}^+\rightarrow){l}^+ \tilde{\chi}_1^0$ and $ l^+
(\tilde{l}^-\rightarrow) {l'}^- \tilde{\chi}_1^0$.  The invariant
mass ($m_{l'l}$) distribution of the leptons is given as
\begin{eqnarray}
\frac{d \Gamma}{d m_{l'l}^2} 
&=&
\sum_{X,Y} \int_{q^2\ge0} d q^2 \rho(q^2) A_{XY}(q^2) 
\nonumber\\
&&\left(
\Gamma^{(l)}_{XY}(q^2) Br^{(l')\star}_{XY}(q^2) 
+
\Gamma^{(l')\star}_{XY}(q^2) Br^{(l)}_{XY}(q^2) 
\right).
\end{eqnarray}
Here $\Gamma^{(l)}_{XY}(q^2)$, $Br^{(l)}_{XY}(q^2)$,  
and $\rho(q^2)$ are 
\begin{eqnarray}
\Gamma^{(l)}_{XY}(q^2)
&=&
\frac{g_2^2}{32 \pi}\left\{R^{(n)}_{l2X} R^{(n)\star}_{l2Y} +L^{(n)}_{l2X} L^{(n)\star}_{l2Y}\right\}
m_{\tilde{\chi}_2^0}
\left(1-\frac{q^2}{m_{\tilde{\chi}_2^0}^2}\right)^2,
\\
Br^{(l)}_{XY}(q^2)
&=&
\frac{g_2^2}{16 \pi}\left\{R^{(n)}_{l1X} R^{(n)\star}_{l1Y} +L^{(n)}_{l1X} L^{(n)\star}_{l1Y}\right\}
\frac{q^2}{[m\Gamma]_{\tilde{l}_X\tilde{l}_Y}}
\left(1-\frac{m_{\tilde{\chi}_1^0}^2}{q^2}\right)^2,
\\
\rho(q^2) &=&
\left\{ 
\begin{array}{ll} 
 \frac{1}{(m_{ll}^{\rm max}(q^2))^2}&(0\le m_{l'l}^2 \le (m_{ll}^{\rm max}(q^2))^2)
\\
 0&(m_{l'l}^2 > (m_{ll}^{\rm max}(q^2))^2)
\end{array}
.\right.
\end{eqnarray}
where $[m\Gamma]_{XY}= (m_{\tilde{l}_X}\Gamma_{\tilde{l}_X} +
m_{\tilde{l}_Y}\Gamma_{\tilde{l}_Y})/2$, and 
\begin{eqnarray}
(m_{ll}^{\rm max}(q^2))^2&=&
m_{\tilde{\chi}_2^0}^2(1- \frac{q^2}{m_{\tilde{\chi}_2^0}^2})
(1-\frac{m_{\tilde{\chi}_1^0}^2}{q^2}).
\end{eqnarray}
 The interaction Lagrangian of fermion-sfermion-neutralino is written
as
\begin{equation}
  {\cal L}_{\rm int}
=  -g_2 \overline{\tilde \chi^0_A} (R^{(n)}_{iAX} P_R +L^{(n)}_{iAX} P_L) l_i 
    \tilde{l}_X^\dagger +h.c.,
\end{equation}
and the coefficients are
\begin{eqnarray}
  L_{iAX}^{(n)}&=& \frac{1}{\sqrt{2}} \{
       [-(O_N)_{A2} -(O_N)_{A1} t_W] U_{X,i}
        + \frac{m_{l_i}}{m_W \cos \beta} (O_N)_{A3} U_{X,i+3} \},
\nonumber \\
  R_{iAX}^{(n)} &=& \frac{1}{\sqrt{2}} \{ 
           \frac{m_{l_i}}{m_W \cos \beta} (O_N)_{A3} U_{X,i} 
           +2 (O_N)_{A1} t_W U_{X,i+3} \}.
\end{eqnarray}
The function of slepton momentum $A_{XY}(q^2)$ is 
\begin{eqnarray}
A_{XY}(q^2) &=& \frac{1}{1+i x_{XY}} 
\frac{\delta(q^2 - m_{\tilde{l}_X}^2) +\delta(q^2 - m_{\tilde{l}_Y}^2) }{2}
\end{eqnarray}
with $x_{XY} = (m_{\tilde{l}_X}^2 -m_{\tilde{l}_Y}^2)/2 [m\Gamma]_{XY}$.

We can simplify above formula assuming two-flavor mixing of the
right-handed slepton and the almost degenerate masses,
\begin{eqnarray}
\frac{d \Gamma}{d m_{l'l}^2}
&=&
\frac{2 \Gamma_0 Br_0}{(m^{\rm max}_{ll}(\overline{m}^2_{\tilde{l}}))^2}
\frac{x^2_{l'l}}{2(1 + x^2_{l'l})} \sin^2 2 \theta
\end{eqnarray}
for 
$0 \le m_{l'l}^2 \le (m_{ll}^{\rm max}(\overline{m}^2_{\tilde{l}}))^2$. 
Here,
\begin{eqnarray}
\Gamma_0
&=&
\frac{g_Y^2}{16 \pi} 
m_{\tilde{\chi}^0_2}
\left(1-\frac{\overline{m}_{\tilde{l}}^2}{m_{\tilde{\chi}_2^0}^2}
\right)^2 [O_N]_{21}^2,
\\
Br_0
&=&
\frac{g_Y^2}{8 \pi}
\frac{\overline{m}_{\tilde{l}}^2}{[m\Gamma]_{\tilde{l}_X\tilde{l}_Y}}
\left(1-\frac{m_{\tilde{\chi}_1^0}^2}{\overline{m}_{\tilde{l}}^2}\right)^2
[O_N]_{11}^2
,
\end{eqnarray}
and $\overline{m}^2_{\tilde{l}}$ and $\sin 2 \theta$ are the average mass and the 
mixing angle for the sleptons.

Next, we present formula for the LFV lepton decays $\mu\rightarrow e
\gamma$ or $\tau\rightarrow\mu\gamma$. Those rates are also written in
similar Lagrangian though, this time, contributions from chargino
loops are also important. The Lagrangian involving
chargino-slepton-lepton is given as
\begin{eqnarray}
{\cal L} &=& -g_2 \bar{l}(L^{(c)}_{iAX} P_R +
R^{(c)}_{iAX} P_L)\tilde{\chi}^-_A \tilde{\nu}_X
\end{eqnarray} 
where $\tilde{\chi}^-_A(A=1,2)$  is a chargino mass eigenstate. 
The coefficients are 
\begin{eqnarray}
 L^{(c)}_{iAX}& =& (O_R)_{A1} U^{\nu}_{X,i},
\nonumber \\
 R^{(c)}_{iAX}& = & \frac{m_{l_i}}{\sqrt{2}m_W\cos\beta}(O_L)_{A2}
                    U^{\nu}_{X,i}.
\end{eqnarray}
Then 
\begin{equation}
\Gamma(l_j \rightarrow l_i \gamma) = \frac{e^2}{16\pi}m^5_{l_j}
(\vert A^L \vert ^2 +\vert A^R\vert ^2)
\end{equation}
where
\begin{equation}
A^R= A^{(n)R} + A^{(c)R}, \ \  A^L= A^{(n)L} + A^{(c)L}.  
\end{equation}
The coefficients in above equations are given as \cite{Hisano:1995cp}
\begin{eqnarray}
A^{(n)R} &=& \frac{1}{32\pi^2}\frac{1}{m^2_{\tilde{l}_X}}
\left[ R^{(n)}_{iAX}R^{(n)\star}_{iAX}\frac{1}{6(1-x_{AX})^4}
\right.
\cr
&&\left.
\times (1-6\xax+ 3 \xax ^2 + 2 \xax ^3 -6\xax ^2\ln \xax)
\right.
\cr
&&\left.
+ R^{(n)}_{iAX} L^{(n)\star}_{jAX} 
\frac{m_{\neu_A}}{m_{l_j}}
\frac{1}{(1-\xax) ^3}
(1-\xax ^2+ 2\xax\ln\xax)\right],
\cr
A^{(c)R}&=&-\frac{1}{32\pi^2}\frac{1}{m^2_{\tilde{\nu}_X}}
\left[ R^{(c)}_{iAX}R^{(c)\star}_{jAX} \frac{1}{6(1-\xax)^4}
\right.\cr
&&\left.
\times (2+ 3\xax - 5 \xax^2 + \xax ^3+ 6\xax\ln \xax)\right.
\cr
&&\left. +R^{(c)}_{iAX}L^{(c)\star}_{jAX}\frac{m_{\cha_A}}{m_{lj}}
\frac{1}{(1-\xax)^2}(-3+4\xax-\xax ^2-2\ln\xax)\right],\cr
&& A^{(n)L}= A^{(n)R}\vert_{L\leftrightarrow R },\ \ 
A^{(c)L}= A^{(c)R}\vert_{L\leftrightarrow R }.
\end{eqnarray}

\section*{B. Sample points}
\setcounter{equation}{0}
\renewcommand{\theequation}{B.\arabic{equation}} In this paper, we
used MC simulation data for two sample points to estimate the event
distribution and the acceptance. We summarize the masses of SUSY
particles and decay branching ratios here because they depends on
choice of gauge couplings and so on.

First, we list mass parameters and relevant SUSY-particle masses in
GeV for the point I), which is studied in this paper. The ISAJET
\cite{Baer:1999sp} is used to generate this spectrum.
\begin{center}
\begin{tabular}{|c|c||c|c||c|c||c|c|}\hline
 $m$  & 100.0  & $\tan\beta$ & 2.1 &  &  & &  \\\hline
 $M$ & 300  &  $\mu$& 497.87 &
 $M_1$ & 126.23  &  $M_2$& 252.36\\ \hline
\hline
 $\mer$ & 157.21  & $\mel$ & 238.78  & $\mupl$ & 654.11 &
 $\mdnl$ & 657.18 \\ \hline
 $\msnu$ & 230.19  &  $\mlsp$ & 121.52  &$\mupr$  
& 630.95 & $\mdnr$ & 628.38\\ \hline
 $\mzii$ & 233.02  & 
 $\mziii$ & 499.18  &$\mwi$ & 232.03 & $\mwii$ & 520.11\\ \hline
 $\mziv$ & 523.23  & $\mntau$ &230.14 &
 $\mtl$ & 459.55 & $\mth$ & 670.68\\ \hline
 $\mtaul$ & 156.81 & $\mtauh$ & 238.92 &
 $\mbl$ & 600.30 & $\mbh$ & 628.84 \\ \hline
 $m_h$ & 94.33 & $m_P$ & 606.79 &  $m_H$& 611.67& $m_{\tilde{g}}$& 732.94 
\\ \hline
\end{tabular}
\end{center}

For the parameter gluino could decay into squarks, and especially, 
it has enhanced branching ratios to the third-generation SUSY particles. 
The first-generation SUSY particles would be also  generated and decay into
chargino or neutralinos to produce leptons. 
The major branching ratios in (\%) are following;

\begin{center}
\begin{tabular}{|l|c||l|c|}
\hline
$ \gl   \rightarrow  \sbi    b$ & 15 &
$ \gl   \rightarrow  \sbii   b$ & 9.7 
\cr \hline
$ \gl   \rightarrow  \sti   t$  & 15 &
$\sul  \rightarrow  \zii  u $&  33
\cr \hline
$\sul \rightarrow  \wi  d $&          65 &
$\sdl \rightarrow  \zii d $&           31
\cr \hline 
$\sdl  \rightarrow  \wi  u $&          64 &
$ \sti \rightarrow  \lsp   t $ &  23
\cr \hline
$ \sti \rightarrow  \zii  t $   &  12 &
$ \sti \rightarrow  \wi  b $    &  64
\cr \hline
$ \sbi \rightarrow  \zii  b   $&      28 &
$ \sbi  \rightarrow \wi  t    $&      41
\cr \hline
$ \sbi   \rightarrow \sti W^-    $&  27 &
$ \sbii \rightarrow  \lsp   b  $   &       67
\cr \hline
$ \sbii \rightarrow  \wi  t  $     &        9.6 &
$ \sbii  \rightarrow  \sti W^-  $   &      15
\cr\hline
$ \zii   \rightarrow  \ser  e $&  9.2 &
$ \zii   \rightarrow  \tilde{\mu}_R  \mu $& 9.2
\cr \hline
$ \zii   \rightarrow  \stau_1 \tau $&12.1 &
$ \wi  \rightarrow  \lsp   W^-   $& 89.7
\cr \hline
$ \wi  \rightarrow  \snue   e  $&  1.4 &
$ \wi  \rightarrow  \snut   \tau $& 0.5
\cr  \hline
$ \wi  \rightarrow  \staul  \nu_{\tau}$ &  6.6 &
&
\cr\hline
\end{tabular}
\end{center}

For point II), the input mass parameters and the SUSY particle masses
are following;
\begin{center}
\begin{tabular}{|c|c||c|c||c|c||c|c|}\hline
 $m_{16}$  & 90.0  & $\tan\beta$ & 10 & &&& \\\hline
 $M$ & 250  &  $\mu$& 199.85 &
 $M_1$ & 103.9  &  $M_2$& 208.75\\ \hline
\hline
 $\mer$ & 139.3  & $\mel$ & 206.09  & $\mupl$ &556.07 &
 $\mdnl$ & 561.67 \\ \hline
 $\msnu$ & 190.28  &  $\mlsp$ & 93.18  &$\mupr$  
& 534.36 & $\mdnr$ & 533.21\\ \hline
 $\mzii$ & 155.13  & 
 $\mziii$ & 208.74  &$\mwi$ & 148.44 & $\mwii$ & 272.52\\ \hline
 $\mziv$ & 273.8  & $\mntau$ &188.67 &
 $\mtl$ & 374.43 & $\mth$ & 563.81\\ \hline
 $\mtaul$ & 132.56 & $\mtauh$ & 206.06 &
 $\mbl$ & 498.27 & $\mbh$ & 531.2 \\ \hline
 $m_h$ & 112.59 & $m_P$ & 436.98 &  $m_H$& 437.63& $m_{\tilde{g}}$& 624.36 
\\ \hline
\end{tabular}
\end{center}
Relevant branching ratios in \% for the sample 
parameter are 
\begin{center}
\begin{tabular}{|c|c||c|c|}
\hline
$ \gl   \rightarrow  \sbi  b $  &  17.4  & 
$ \gl   \rightarrow  \sbii b $  &  10.1 
\cr \hline
$ \gl   \rightarrow  \sti t $  &  13.2  &
$\sul\rightarrow\zii  u$  & 20.7 
\cr \hline 
$\sul\rightarrow\ziii u$  & 0.4  & 
$\sul\rightarrow\ziv  u$  & 12.2 
\cr \hline 
$\sul\rightarrow\wii  d$  & 21.4 & 
$\sdl\rightarrow\zii  d$ &14.5  
\cr \hline
$\sdl\rightarrow\ziii d$ & 0.7  &
$\sdl\rightarrow\ziv  d$ & 15.4 
\cr \hline 
$\sdl\rightarrow\tilde{\chi}^+_2 u$ & 36.4 &
$ \sti    \rightarrow  \lsp  t $ & 9.5 
\cr \hline
$ \sti    \rightarrow  \zii t $ & 10.1 &
$ \sti    \rightarrow  \wi  b $ & 74.3
\cr \hline
$ \sti    \rightarrow  \wii  b $ & 6.1 &
$ \sbi \rightarrow  \tilde{\chi}^+_1  t     $  &   29.9
\cr \hline
$ \sbi \rightarrow  \tilde{\chi}^+_2  t     $  &   37.8 &
$ \sbi \rightarrow  \zii  t    $  &    9.8
\cr \hline
$ \sbi \rightarrow  \sti W^- $  &   10.0 &
$ \sbii \rightarrow  \tilde{\chi}^+_1  t    $  &   14.1
\cr \hline
$ \sbii \rightarrow  \tilde{\chi}^+_2  t    $  &   35.5 &
$ \sbii \rightarrow  \tilde{\chi}^2_0  b   $  &    4.9
\cr \hline
$ \sbii \rightarrow  \sti W^-$  &   10.5 &
$\ziv\rightarrow\sel e$& 5.5 
\cr \hline 
$\ziv\rightarrow\ser e$& 1.1 & 
$\zii\rightarrow\ser e$ & 23.6 
\cr \hline
$\wii\rightarrow\snu e$ & 9.7  & 
$\wi\rightarrow\staul\nu_\tau$ & 89.1 
\cr \hline
$\snu\rightarrow\wi e$ & 40.1 &
$\sel\rightarrow\zii e$ & 38.2 
\cr \hline
$\sel\rightarrow\lsp e$ & 19.8 & 
$\ser\rightarrow\lsp e$ &100  
\cr \hline
\end{tabular}
\end{center}

\end{document}